\documentclass[fleqn,10pt]{wlscirep}
\usepackage[utf8]{inputenc}
\usepackage[T1]{fontenc}
\usepackage{cancel}
\usepackage[normalem]{ulem} 
\usepackage{soul} 
\title{Quantum control and characterization of ultrafast ionization with orthogonal two-color laser pulses}

\author[1,*]{Hicham Agueny}

\affil[1]{Department of Physics and Technology, Allegt. 55,
University of Bergen, N-5007 Bergen, Norway}

\affil[*]{hicham.agueny@uib.no}

\begin{abstract}
We study ultrafast ionization dynamics using orthogonally polarized two-color (OTC) laser pulses involving the resonant "first plus second" ($\omega+2\omega$) scheme. The scheme is illustrated by numerical simulations of the time-dependent Schr\"odinger equation and recording the photoelectron momentum distribution. On the basis of the simulations of this resonant ionization, we identify signatures of the dynamic Autler-Townes effect and dynamic interference, in which their characterization is not possible in spectral domain. Taking advantage of the OTC scheme we show that these dynamical effects, which occur at the same time scale, can be characterized in momentum space by controlling the spatial quantum interference. In particular, we show that with the use of this control scheme, one can tailor the properties of the control pulse to lead to enhancement of the ionization rate through the Autler-Townes effect without affecting the dynamic interference. This enhancement is shown to result from constructive interferences between partial photoelectron waves having opposite-parity, and found to manifest by symmetry-breaking of the momentum distribution. The scenario is investigated for a prototype of a hydrogen atom and is broadly applicable to other systems. Our findings may have applications for photoelectron interferometers to control the electron dynamics in time and space, and for accurate temporal characterization of attosecond pulses.
\end{abstract}
\begin{document}

\flushbottom
\maketitle
%
%
\thispagestyle{empty}


\section*{Introduction}

Developments in ultrafast electron diffraction and microscopy have made it possible to follow dynamical changes in the course of atomic and molecular transitions in time and space \cite{Zewail:10}. Their coherent control was achieved by tailoring the properties of laser pulses and recording structural changes in the measured momentum distribution \cite{Huismans:11,Bian:12}. The latter carries valuable information about the target structure and encodes the dynamics of atoms.   

Recently, there has been a great interest in using orthogonally polarized two-color (OTC) laser pulses to gain control over the electron dynamics. This has been shown to offer a new and intriguing degree of control freedom for light-matter interactions. The OTC scheme has been used, for instance, to control temporal double-slit interference \cite{Richter:15}, to disentangle intracycle interference  \cite{Xie:17}, to identify the nonadiabatic subcycle electron dynamics \cite{Geng:15}, and to study the effect of Coulomb focusing \cite{Richter:16}. Measuring the electron correlation in double ionization process by OTC fields was also reported \cite{Zhang:14,Zhou:11}. All these phenomena have been observed in the photoelectron momentum distributions. Measuring these distributions has the advantage of providing more information than the angle-integrated spectra, as has been reported in many theoretical and experimental studies \cite{Eckle:08,Pfeiffer:12a,Pfeiffer:12b,Xie:15}. Time-resolved photoelectron holography is an example that reveals temporal and spatial interferences between partial waves, thus allowing to reconstruct holographic images of the electron wavepacket \cite{Villeneuve:17}.

Dynamic interference is another class of temporal interferences: It was shown to arise from the interference between electrons released at the rising and falling edges of a pulse \cite{Toyota:07,Demekhin:12a}, and manifest by a splitting of the above-threshold ionization peaks. The observation of this dynamic interference, in particular in the work of Demekhin \textit{et al.} \cite{Demekhin:12a}, has sparked off further theoretical efforts to understand the complete
nature of the phenomenon \cite{Demekhin:13,Yu:13,Lars:14,Robicheaux:16,Morten:16,Agueny:17a,Rost:17}. It was, however, shown that the obtained results in \cite{Demekhin:12a}, which were discussed in connection with one-photon ionization from the hydrogen ground state, are invalid \cite{Rost:17}. On the other hand, the emergence of the phenomenon has been reported for a particular case, where the atomic potential supports only one or two bound states  \cite{Toyota:07,Toyota:08,Demekhin:12b,Rost:18}. 

A more complex situation is, however, when atomic transitions mediated by a coherent resonant laser pulse are involved in the dynamics. Here, the well-known Autler-Townes effect \cite{Autler:55} emerges in the spectral domain as a manifestation of Rabi-oscillations, and thus any signature of dynamic interference should be disentengled from the above-mentioned effect. In this resonant dynamics, it was found that, when the electric field is strong enough to induce many Rabi-cycles, additional sub-peaks were observed in the angle-integrated spectrum accompanied with the Autler-Townes doublet \cite{Lewenstein:86,LaGattuta:93,Bauer:07,Demekhin:12b}. Here, most theoretical works have focused on discussing only the Autler-Townes doublet (see e.g. \cite{LaGattuta:93,Bauer:07,Piraux:11,Agueny:19} and references therein), while the origin of the additional sub-peaks has not received much attention except for some theoretical works \cite{Lewenstein:86,Rongqing:91,Bauer:07}, where some suggestions have been evoked either in terms of temporal interference effects \cite{Lewenstein:86}, or high-order ionization processes \cite{Rongqing:91,Bauer:07}. On the other hand, although dynamic interference has been reported for a resonant two-photon ionization process \cite{Demekhin:12b}, disentangling this phenomenon from the Autler-Townes effect has not been emphasized. 

Moreover, this class of interferences was investigated mostly by examining the angle-integrated photoelectron spectrum. Here, interesting effects related to quantum coherence due to the opposite-parity of the emitted electrons might be washed out. This however, can be observed by measuring the photoelectron momentum distribution, which permits to resolve dynamical signals that can be separated with the use of an OTC scheme \cite{Xie:17}. Indeed, during the photoionization process, the outgoing photoelectron wave packet is partitioned in a form of a coherent superposition of partial photoelectron waves with a different parity. When the pulse envelope supports many cycles, dynamic interference might occur between partial waves having the same parity and can be identified in the spectral domain, while the interference of those with opposite-parity can be recorded in the angular distribution. Providing insights into the interplay between these temporal and spatial interferences requires to carry out the photoelectron momentum distribution, where the footprint of the coherent control of the electron dynamics can be mapped out.

Note that most of the studies of coherent control of the photoelectron momentum distribution by means of a two-color scheme have focused on nonresonant photoionization processes with wavelengths mostly in the visible (Vis) and infrared (IR) ranges  (see, e.g.,  \cite{Richter:15,Xie:17,Geng:15}). Involving resonances in the dynamics is of particular interest for quantum coherent control \cite{Gruson:16,Kaldun:16}. For instance, it has been shown that the presence of a resonant intermediate state modifies dramatically the photoionization spectra \cite{Martin:14} and angular distributions \cite{Ishikawa:12,Agueny:17,Agueny:18b} as a result of the interference between resonant and nonresonant paths involved during the excitation \cite{Agueny:17,Agueny:18b} and ionization \cite{Martin:14,Ishikawa:12} processes. The role of resonances has also been discussed in connection with quantum coherence and found to be an essential criterion for increasing the degree of coherence \cite{Carlstrom:18}. Furthermore, coherent control experiments with shorter wavelengths than those in the Vis and IR ranges was demonstrated \cite{Prince:16}, which promises to open up a new field of research \cite{Hartmann:16,Giannessi:18}.

In this work, we investigate the resonant two-photon ionization process of a hydrogen atom complementing the well-known effects that emerge during this process and providing insight into the mechanism underlying the additional sub-structures observed in the spectral domain. This is achieved by presenting a detailed study on the effect of a control laser pulse within an OTC scheme in strong-field ionization. In the proposed two-color scheme, the ionization dynamics is mainly induced by the fundamental harmonic field, which is chosen to be resonant with the $1s-2p$ transition, while its second harmonic is used to control the induced dynamics. The time-propagation scheme is performed  by solving the two-dimensional time-dependent Schr\"odinguer equations  (2D-TDSE). 

Taking advantage of the OTC scheme in being selectively sensitive to the generation of partial photoelectron waves, we show that the use of this scheme allows to characterize signals originating from electrons that follow the pulse envelope and those that are driven by the instantaneous electric field in momentum space. In particular, we show that the above-mentioned sub-structures are caused by dynamic interference, and which its signature can be disentangled from that of the Autler-Townes effect by carrying out the momentum distribution. We further show that the proposed scheme allows to coherently control the resonant process by exploiting the Autler-Townes effect. The control scheme relies on quantum-path interferences involved in the photoionization processes. By tuning the properties of the control scheme, we show that the ionization rate can be enhanced owing to the constructive interference between multiple photoionization pathways. As a result, intriguing changes in the photoelectron momentum distributions are recorded, and are found to manifest by symmetry-breaking and splitting of the localised momentum distribution. We show that the emergence of these effects can be manipulated by varying the relative optical carrier envelope phase and the relative intensity ratio of the two pulses. This offers the potential for controlling the continuum wave packet, which is encoded in the interference fringes.


\section*{Results}


\subsection*{One-colour scheme}



 

We first consider the one-colour scheme which consists of the fundamental harmonic that is resonant with the $1s-2p$ transition. This resonant two-photon ionization process has been discussed extensively and mostly in connection with the photoelectron spectrum \cite{Lewenstein:86,LaGattuta:93,Bauer:07,Piraux:11,Demekhin:12b}.  
\begin{figure*} [h!]
\centering
\includegraphics[width=14.cm,height=12.0cm]{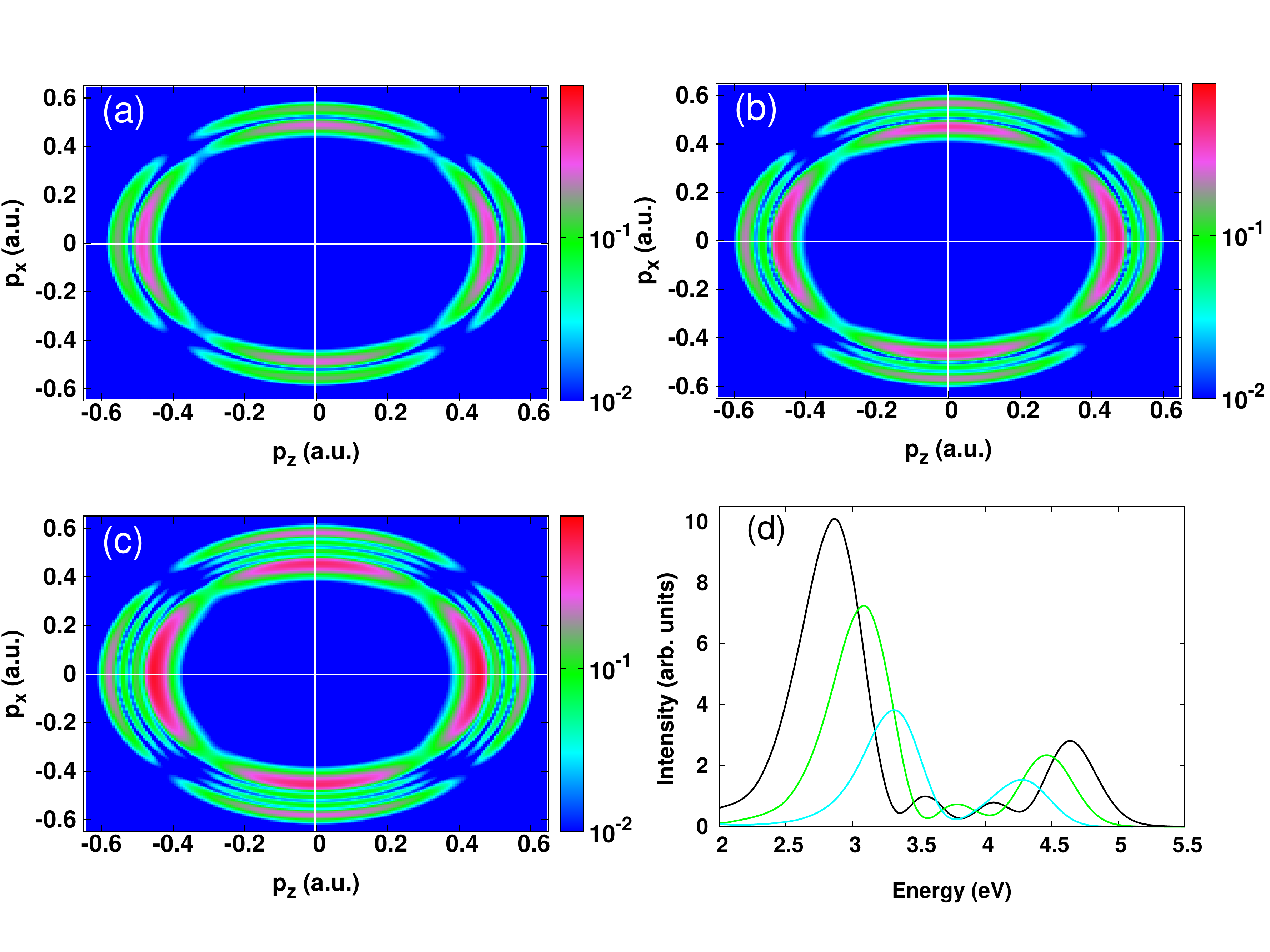}
\caption{\label{fig1} One-color scheme calculations. (a)-(c) Photoelectron momentum
distributions in the $p_x-p_z$ plane. (d) Angle-integrated photoelectron spectrum. The fundamental harmonic field having the angular frequency $\omega$=8.58 eV ($\lambda_1$=144.53 nm) is linearly polarized along the $z$ axis and the optical phase is zero. The results are shown for three peak intensities: (a) 6 $\times$10$^{13}$ W/cm$^2$ (cyan curve); (b) 1.2 $\times$10$^{14}$ W/cm$^2$ (green curve); (c) 2 $\times$10$^{14}$ W/cm$^2$ (black curve).}
\end{figure*}
At this resonant frequency of a strong electric field, Rabi flopping is expected to emerge in the occupation probabilities oscillating between the states $1s$ and $2p$ in time domain. In the energy domain, it manifests by the Autler-Townes effect (or also called AC Stark splitting), which leads to a doublet-peak splitting of the photoelectron spectrum. 

In Fig. ~\ref{fig1}  we summarise the results stemming from our 2D-model for ultrafast ionization dynamics from a hydrogen atom using a resonant harmonic field (the corresponding photon energy is $\omega_1=$ 8.58 eV and the pulse duration is $\tau_1$=638 a.u. =15.4 fs). We stress here that the resonant frequency employed in this model is different from the exact one which is 10.2 eV. This difference originates from the use of a soft potential model in equation (\ref{eq2}), which leads to an energy of -5 .0274 eV (-0.1847 a.u.) for the first excited state, while the exact one is -3.4014 eV (-0.125 a.u.).  We show however that this difference in energy does not affect the physics involved in the dynamics and which is discussed below. This is illustrated by solving the TDSE in three-dimension, thus validating the results obtained within our 2D-model (see the Supplemental Material \cite{comment} for details about the calculations and the momentum distributions.) 

The photoelectron momentum distributions obtained with the 2D-model are shown in Fig. ~\ref{fig1} at three-peak intensities: 6 $\times$10$^{13}$ (Fig. ~\ref{fig1}(a)), 1.2 $\times$10$^{14}$ (Fig. ~\ref{fig1}(b)) and 2 $\times$10$^{14}$ W/cm$^2$ (Fig. ~\ref{fig1}(c)), where two, three and four Rabi-cycles are completed, respectively. It is seen that the momentum distributions exhibit rings, signature of the splitting of the localised distributions. The splitting behavior is strongly modified when varying the pulse intensity; in particular, increasing the peak intensity results in additional rings, whose number is determined by the number of completed Rabi-cycles. Here, the Autler-Townes doublet is expected to be localized at the momentum $\sqrt {2(2\omega_1-I_p \pm \Omega_{Rabi}/2)}$ , where $I_p$ is the ionization potential and $\Omega_{Rabi}=E_0\langle1s \mid  z \mid 2p\rangle$ is the Rabi frequency.  

Further details about the splitting and particularly the Autler-Townes doublet are provided by the angle-integrated photoelectron spectra as displayed in Fig. ~\ref{fig1}(d). The Autler-Townes effect should manifest by two pronounced peaks, since its origin arises from the states which are resonantly coupled to the continuum \cite{Autler:55,Knight:79,Bauer:07}. These two peaks can be clearly identified in the photoelectron spectra as shown in Fig. ~\ref{fig1}(d). The numerical values of these peak positions extracted from the photoelectron spectra are summarised in Table ~\ref{table1} and compared to those obtained from the well-known analytical formula $2\omega_1-I_p \pm \Omega_{Rabi}/2$. Here the sign (-) refers to the left Autler-Townes peak and (+) refers to the right one. Also is shown the energy spacing between the left and the right peaks and is compared to the expected Rabi-frequency $\Omega_{Rabi}$. Here, the energy separation is seen to increase with increasing the peak intensity; in accordance with the analytical formula of the Rabi-frequency $\Omega_{Rabi}$. The numerical values of the energy separation are in general close to the expected Rabi-frequency, although an energy difference of 0.1 up 0.5 eV is noticed depending on the peak intensity. A similar discrepancy was observed in \cite{Bauer:07} and discussed in terms of the dynamic AC Stark effect of the ground and excited states. This effect was found to lead to a shift of the position of the Autler-Townes peaks \cite{Bauer:07}. The good agreement between the numerical values predicted by the analytical formula and those obtained from the TDSE calculations suggests that these two pronounced peaks can be attributed to the Autler-Townes effect.

\begin{table}[h]
\centering
\caption{Comparison of the Autler-Townes (AT) peaks extracted from the photoelectron spectra (cf. Fig. ~\ref{fig1}(d)) based on TDSE and those from the analytical formula : $2\omega_1-I_p \pm \Omega_{Rabi}/2$. Here the sign (-) refers to the left AT peak "Left" and (+) refers to the right one "Right". The energy spacing between these two peaks is also shown and compared to the analytical $\Omega_{Rabi}$. The comparison is made for three peak intensities: 6 $\times$10$^{13}$, 1.2 $\times$10$^{14}$ and 2$\times$ 10$^{14}$ W/cm$^2$. The angular frequency is fixed at $\omega_1$=8.5784 eV. \\}
\label{table1}

\begin{tabular}{ c|c|c|c||c|c|c| } 
\cline{2-7}
& \multicolumn{3}{ c ||}{TDSE} & \multicolumn{3}{ c |}{Analytic} \\
\hline
 & \multicolumn{2}{ c |}{AT doublet}& $\Omega_{Rabi}$ &  \multicolumn{2}{ c |}{AT doublet} & $\Omega_{Rabi}$  \\
  {Laser intensity (W/cm$^2$) } & \multicolumn{2}{ c |}{(eV)}& (eV )&  \multicolumn{2}{ c |}{(eV)} & (eV) \\
& Left & Right &  & Left & Right & \\
\hline\hline
6.0 10$^{13}$ & 3.313 & 4.474 & 1.161 & 2.945 & 4.201 & 1.256  \\ 
1.2 10$^{14}$ & 3.087 & 4.464 & 1.377 & 2.685 & 4.461 & 1.776 \\ 
2.0 10$^{14}$ & 2.868 & 4.636 & 1.768 & 2.427 & 4.720 & 2.293 \\ 
\hline
\end{tabular}
\end{table}

\begin{figure*} 
\centering
\includegraphics[width=12.cm,height=6.0cm]{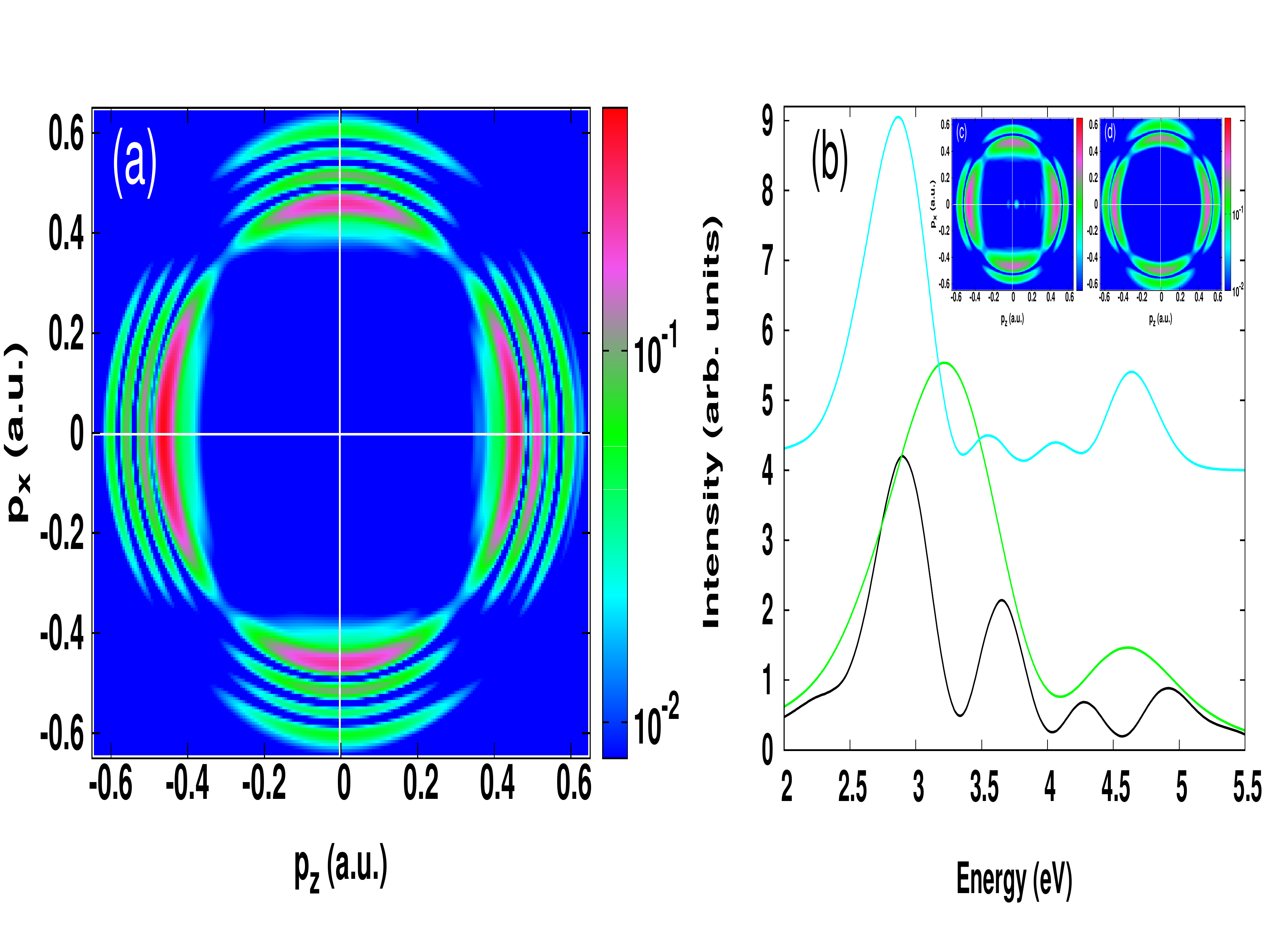}
\caption{\label{fig2} One-color scheme calculations. (a) Photoelectron momentum
distributions in the $p_x-p_z$ plane obtained from a coherent superposition of the ionization wave functions in the rising ($\psi_{rising}(\vec{p})$) and falling ($\psi_{falling}(\vec{p})$) parts of the pulse (i.e. $|\psi_{rising}(\vec{p}) - \psi_{falling}(\vec{p})|^2$).  (b) Angle-integrated photoelectron spectrum obtained from the coherent sum (black curve) and incoherent sum (green curve) (see text). For reference, the exact spectrum (same as in Fig. ~\ref{fig1}(d)) is also shown by a cyan curve with an offset. Inset:  Electron densities in the $p_x-p_z$ plane for the released electrons in the rising (c) $|\psi_{rising}(\vec{p})|^2$ and falling (d) $|\psi_{falling}(\vec{p})|^2$ parts of the pulse.
The results are shown for the peak intensity 2 $\times$10$^{14}$ W/cm$^2$. The angular frequency is fixed at $\omega$=8.58 eV ($\lambda_1$=144.53 nm) and the optical phase is zero.}
\end{figure*}

In Fig. ~\ref{fig1}(d) it is seen that increasing the peak intensity leads to the emergence of additional subpeaks which accompany the Autler-Townes doublet. In the following we show that this additional pattern is related to dynamic interference as arising from the interference of photoelectron wave-packets at the rising and falling edges of the pulse envelope. To this end, we perform additional calculations based on the TDSE, where the ionization wave functions in the rising and falling edges of the pulse are calculated separately. This is particularly carried out for a peak intensity of 2 $\times$10$^{14}$ W/cm$^2$.  In the rising part of the half-pulse, the outgoing wave-packet $\psi_{rising}(\vec{r})$ is obtained by projecting out the dressed bound states, which are obtained at the maximum of the pulse envelope, from the time-dependent wave function. Its density in momentum space is depicted in the inset of Fig. ~\ref{fig2}(b) (left hand-side). 

In the falling part of the half-pulse, the propagation scheme starts from the dressed ground state, which is obtained at the end of the rising part of the pulse envelope. At the end of the falling pulse, the ionization wave function $\psi_{falling}(\vec{r})$ is calculated. The density of the ionized electron in momentum space is displayed in the inset of Fig. ~\ref{fig2}(b) (right-hand side). Here both densities obtained at the rising and falling parts of the pulse show only a pair of rings, signature of the Autler-Townes effect. When a coherent superposition of these two components (i.e. $|\psi_{rising}(\vec{p}) - \psi_{falling}(\vec{p})|^2$) is carried out in momentum space, one sees the emergence of the additional rings (cf. Fig. ~\ref{fig2}(a)); in excellent agreement with the exact results obtained with the full pulse (cf. Fig. ~\ref{fig1}(c)). The angle-integrated photoelectron spectrum stemming from this coherent superposition model reproduces well the oscillatory behavior of the additional patterns in the spectrum using the full pulse, although a slight energy shift of the pattern is seen to emerge. On the other hand, the incoherent sum of these two components of the pulse results only to the doublet peaks. The close agreement between the coherent superposition model and the exact calculation using the full pulse demonstrates the origin of the additional patterns in the spectrum in terms of dynamic interference.

In this scenario, the AC Stark effect plays a role of a coherent beam splitter, which splits the initial bound wave-packet coherently during the dynamics. This partitioning wave-packet follows the time-envelope of the laser pulse and gets released at two different times of the pulse in the rising and falling half-pulses, as has been discussed in \cite{Demekhin:12a,Demekhin:12b}. These released electrons reach the continuum with the same final kinetic energy, and therefore interfere constructively and destructively resulting in additional patterns as identified in Fig. ~\ref{fig1} and Fig. ~\ref{fig2}. On the other hand, the dynamic Autler-Townes effect is seen to emerge in the rising as well as in the falling edges of the pulse envelope, which shows that the temporal build-up of this effect is independent of the nature of the pulse edges \cite{Agueny:19}. 

At this point, we conclude that the dynamic Autler-Townes effect and dynamic interference are identified in the spectral domain, and the spectral analysis suggests that their signals can be disentangled by introducing a second laser pulse. This is supported by the fact that, in multi-cycle regime, the momentum distributions calculated within a one-color scheme are insensitive to the change of the optical phase. As a result, the characterization of these dynamical effects is difficult within a one-color scheme. In the following, we show that the use of OTC laser pulses involving the "first plus second" ($\omega+2\omega$) scheme permits to characterize these signals in momentum space.This will be achieved in the following by investigating the sensitivity of these effects to the properties of the control pulse, thus offering the potential of controlling the photoelectron angular distribution.

\begin{figure*} [h!]
\centering
\includegraphics[width=14.cm,height=12.0cm]{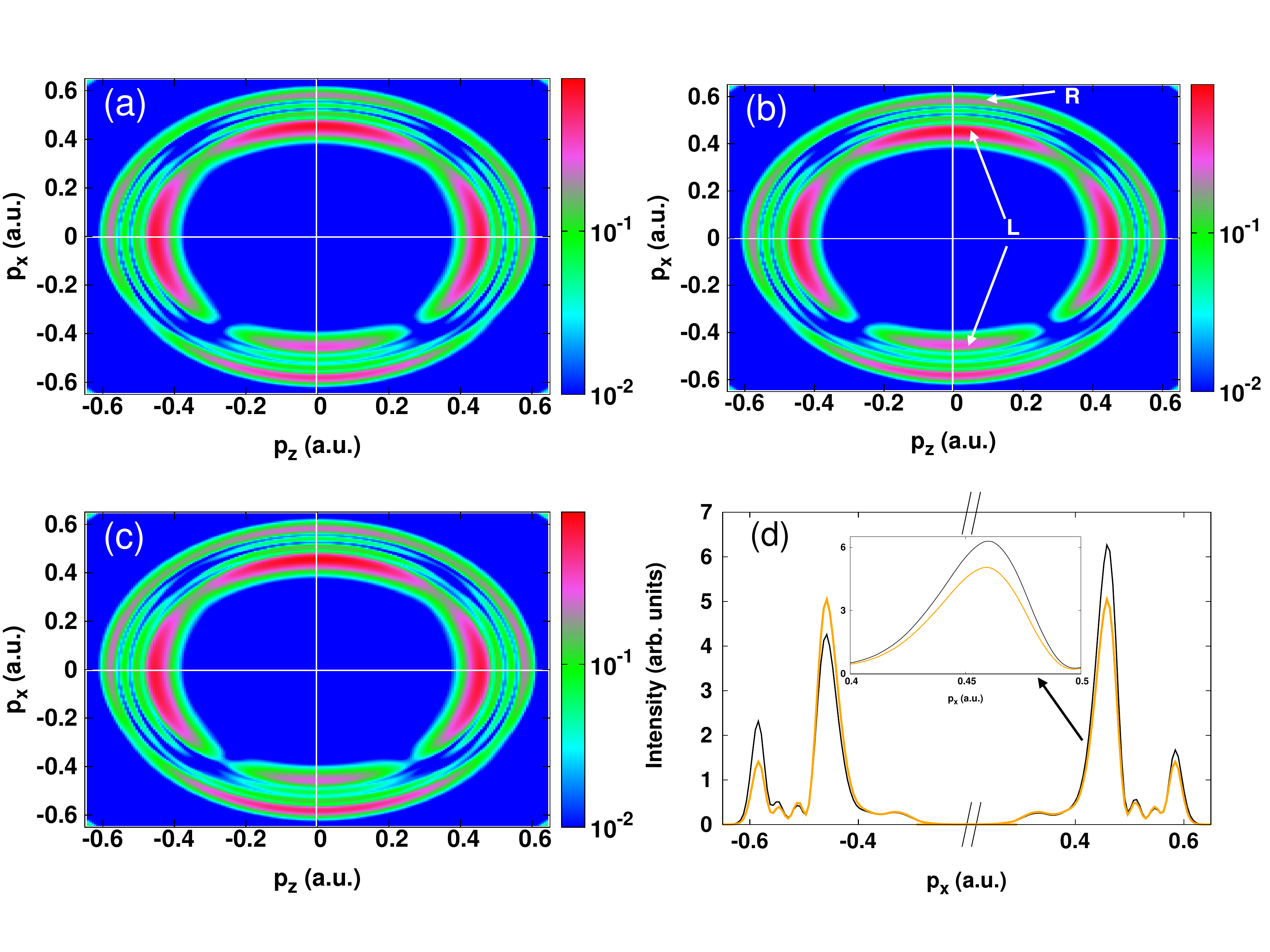}
\caption{\label{fig3} Two-color scheme calculations. (a)-(c) Photoelectron momentum
distributions in the $p_x-p_z$ plane. (d) Angle-integrated photoelectron spectra (integrated over the electron emission angle along the $z$-polarization direction) as a function of $p_x$ at forward $p_x>0$ and backward $p_x<0$ electron emission directions. The fundamental harmonic field is linearly polarized along the $z$ axis and its peak intensity is fixed at 2 $\times$10$^{14}$ W/cm$^2$ and the angular frequency is $\omega$=8.58 eV ($\lambda_1$=144.53 nm). The angular frequency of its second harmonic (streaking field) is $\omega$=17.16 eV ($\lambda_2=\lambda_1/2$=72.26 nm), and the relative optical phase is $\delta\varphi=0$. The results are shown for three relative intensity ratios :
(a) $\eta$=0.45; (b) $\eta$=0.63 (black curve); (c) $\eta$=0.92. Also shown the spectrum obtained within the one-color scheme at the peak intensity of the fundamental field (orange curve). Inset: a zoom of the left Autler-Townes peaks. The white arrows with the letters "L" and "R" indicate the left and right of the Autler-Townes peaks respectively.}
\end{figure*}

\subsection*{Two-colour scheme}

We consider an OTC scheme that comprises a fundamental ($ \lambda_1= 144.53$ nm) pulse and its second harmonic ($ \lambda_2 =\lambda_1/2=$ 72.26 nm), and which are chosen to be linearly polarized, respectively, along $z$ and $x$ axes. The fundamental field contributes mainly to the ionization dynamics and is considered to be resonant with the $1s-2p$ transition (the corresponding photon energy is $\omega_1=$ 8.58 eV), while the role of its second harmonic is determined according to the relative intensity ratio $\eta$. We choose different values of $\eta$ such that the second pulse acts as a control pulse. Its photon energy is fixed at  $\omega_2=2\omega_1=$17.16 eV, and the pulse duration is $\tau_2=\tau_1$ =638 a.u. =15.4 fs. The combined fields are expected to induce interesting effects related to the coherent control of the electron dynamics through the Autler-Townes effect and dynamic interference, which can be realized by varying the relative intensity ratio and optical phase.  

The photoelectron momentum distributions obtained using this OTC scheme are displayed in Fig. ~\ref{fig3} for the peak intensity of the fundamental harmonic 2 $\times$10$^{14}$ W/cm$^2$. The results are shown at the relative optical phase $\delta\varphi=0$ and for three values of the relative intensity ratio: $\eta=$0.45, 0.63 and 0.92. At first glance, the distributions in Fig. ~\ref{fig3} exhibit symmetry-breaking of the momentum distribution and that occurs mainly along the $x$ polarization direction (polarization of the streaking field). This asymmetry exhibits a strong dependence on the relative intensity ratio. A close inspection of these distributions shows that the asymmetry manifests by changes in the shape of the rings. In particular, when comparing the left rings (indicated by white arrows with the letter "L" in Fig. ~\ref{fig3}(b)) of the Autler-Townes doublet located in the lower ($p_x<0$, backward direction) and upper ($p_x>0$, forward direction) parts of the momentum distribution, it is seen that the latter (i.e. the left ring in $p_x>0$ side) gets enhanced and becomes broader, unlike the lower part (i.e. the left ring in $p_x<0$), where the peak becomes less pronounced. On the other hand, the signature of dynamic interference is found to be slightly affected by the change of the relative intensity ratio.  

These observations become clear when looking at the transversal photoelectron spectrum obtained by integrating over the longitudinal part of the electron emission angle, as shown in Fig. ~\ref{fig3}(d). For clarity, we only show the spectrum extracted from Fig. ~\ref{fig3}(b) (i.e for $\eta=$0.63 ) (black curve). We also show in the same figure, for reference, the spectrum obtained from one-color scheme calculations. This helps to highlight the impact of the streaking field on the broadening of the Autler-Townes doublet, which can be seen clearly in the inset of Fig. ~\ref{fig3}(d), where a zoom of the left Autler-Townes peak is depicted. Here, the observed broadening is an interesting feature of the control scheme. It is indeed a signature of the enhancement of the ionization rate due to the presence of the streaking field. This is in agreement with the findings in \cite{Bauer:07,Piraux:11} where the broadening of the left Autler-Townes peak has been shown to arise from the coupling of the dressed states to the continuum (i.e. rate of the ionization). This suggests the possibility of controlling and manipulating the ionization rate through the Autler-Townes effect. 

In the following, we show that the enhancement of the ionization rate results from constructive interferences between one-photon  and two-photon ionization pathways involved in the absorption process in the final state. In other words, we show that this enhancement is a manifestation of the symmetry-breaking of the momentum distribution. 

\begin{figure} [h!]
\centering
\includegraphics[width=6.5cm,height=5cm]{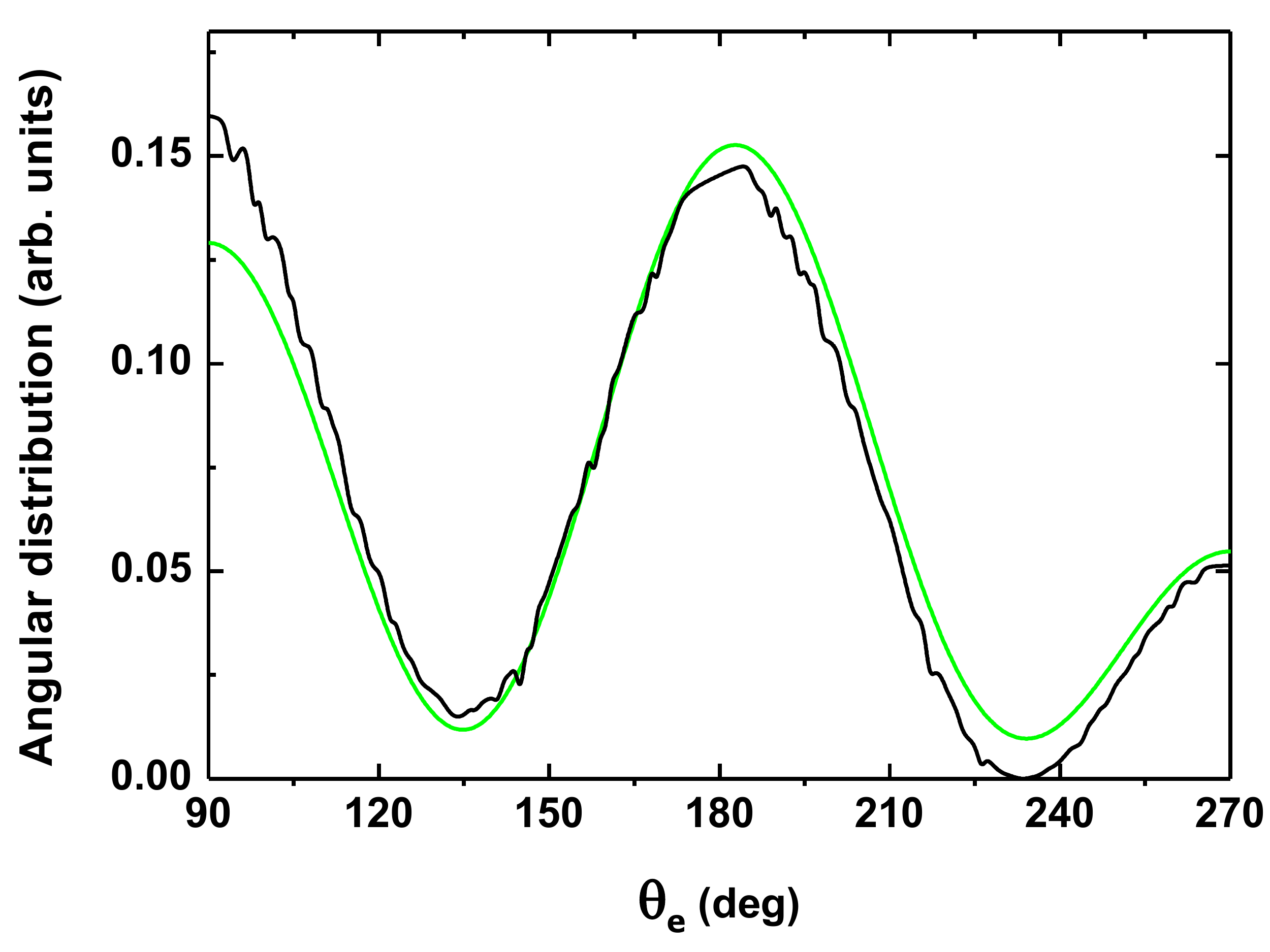}
\caption{\label{fig4} Angular distribution extracted from Fig. 3 (b) at the photoelectron energy $E$=2.87 eV (black curve) and the one obtained from a coherent superposition of three spherical harmonics ($s$-, $p$- and $d$-waves) [cf. Eq. (\ref{eq5})] (green curve). The photoelectron energy corresponds to the left peak of the Autler-Townes doublet. The peak intensity of the fundamental harmonic field is fixed at 2 10$^{14}$ W/cm$^2$ and the angular frequency is $\omega$=8.58 eV ($\lambda_1$=144.53 nm). The electric field is linearly polarized along the $z$ axis. The relative intensity ratio is fixed at $\eta=$0.63 and the angular frequency is $\omega$=17.16 eV ($\lambda_2=\lambda_1/2$=72.26 nm). The relative optical phase is $\delta\varphi$= 0.}
\end{figure}

The starting point is to get a look at the angular distribution of the ejected electron. In Fig. ~\ref{fig4}, we plot the electron angular distribution extracted from the momentum distribution shown in Fig. ~\ref{fig3}(b) at the photoelectron energy $E$=2.87 eV (black curve). The latter energy corresponds to the left peak of Autler-Townes doublet, where the asymmetric ionization along the polarization of the streaking field is enhanced. The obtained result is compared to that stemming from a model based on a coherent superposition of three possible spherical harmonics. This comparison is displayed in the range of the electron ejected angle 90$^o$-270$^o$, where the asymmetric ionization is seen to be enhanced by the streaking field (cf. Fig. ~\ref{fig3}(d)). In this model the total continuum wave function is written as
\begin{equation}\label{ang}
\psi(\theta_e) = \alpha_s Y_{0,0}(\theta_e)e^{i\delta_s}+\alpha_d Y_{2,0}(\theta_e)e^{i\delta_d}+\alpha_p Y_{1,0}(\theta_e -\pi)e^{i\delta_p},
\end{equation}
where $Y_{l,0}(\theta_e)$ ($l$=0,1,2) are spherical harmonics of partial waves of type $s$, $p$ and $d$. $\alpha_i$ and $\delta_i$ ($i$=$s$,$p$, $d$) are respectively, the amplitudes and the corresponding phases of the partial wave contributions. The first two terms of the right side of the equation describe the two-photon pathways ionization induced by the fundamental harmonic generating the $s$- and $d$-waves, whereas the the latter term results from  the direct one-photon ionization by the streaking field producing the $p$-waves. The model in equation (\ref{ang}) permits to extract the amplitude and phase of the partial waves involved in the dynamics, thus allowing to reconstruct the electron wavepacket and manipulate its evolution with the use of a two-color scheme. A similar model has been used by Villeneuve \textit{et al.} \cite{Villeneuve:17}  to fit the experimental data, which has allowed to determine the exact amplitude and phase of each partial wave component. This procedure has enabled to spatially image the angular structure of the continuum wave function \cite{Villeneuve:17}.

Fitting the model in equation (\ref{ang}) to the exact angular distribution (black curve) yields the result shown in Fig. ~\ref{fig4} (green curve). The model reproduces very well the asymmetric profile; thus demonstrating that the interference between partial photoelectron wave packets generated with opposite-parity due to the two components of the OTC fields to be the origin of the observed symmetry-breaking of the momentum distribution. This asymmetry is reflected in the angle-integrated spectra and manifests by the enhancement of the ionization rate. Therefore, the emergence of this asymmetric ionization reveals the role of the streaking field in the OTC scheme as an attractive means for coherent control of the ultrafast ionization process. 

\begin{figure*} [h!]
\centering
\includegraphics[width=14.cm,height=12.0cm]{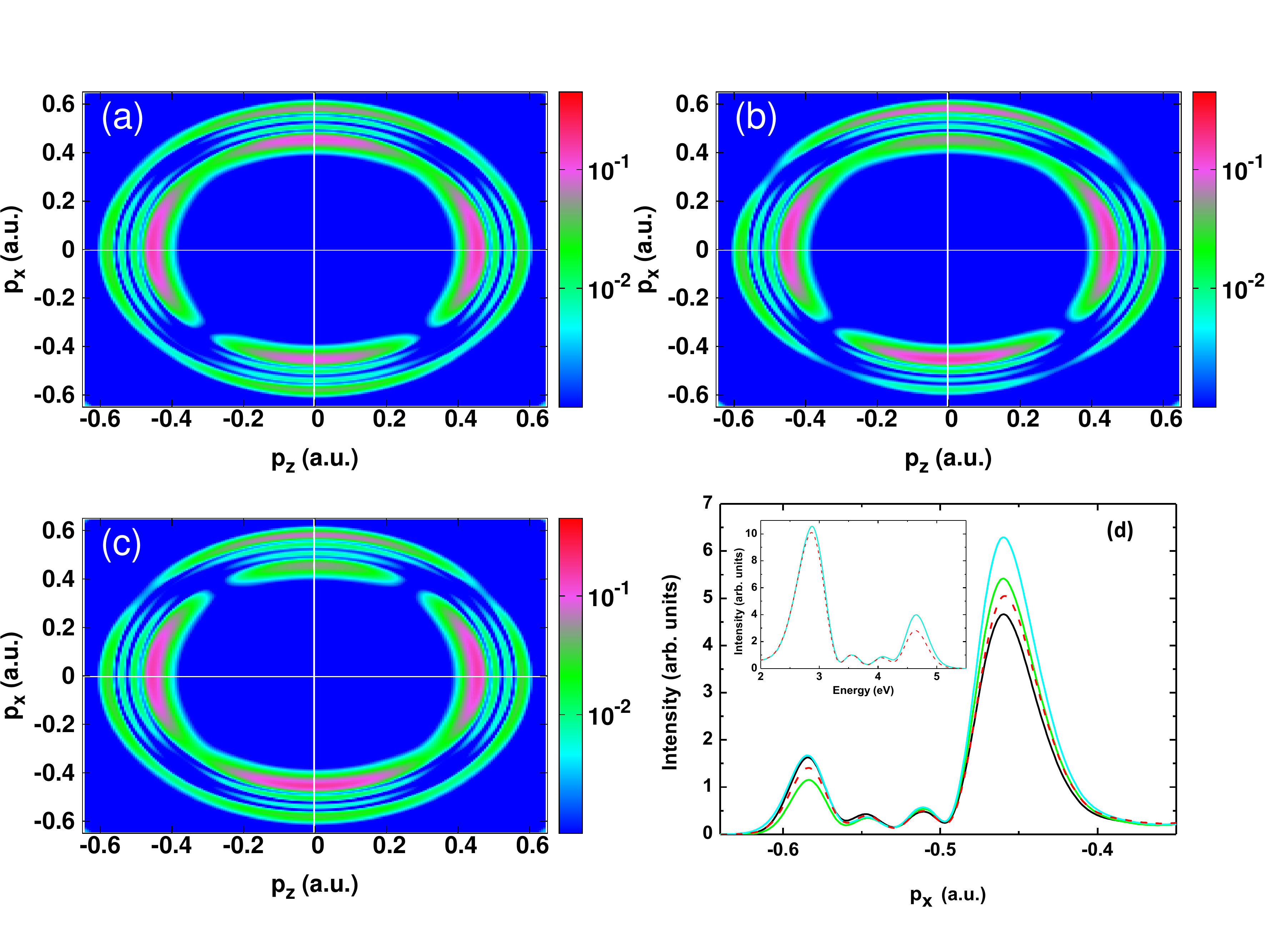}
\caption{\label{fig5} Two-color scheme calculations. (a)-(c) Photoelectron momentum
distributions resulting from the OTC fields in the $p_x-p_z$ plane. (d) Angle-integrated photoelectron spectra (integrated over the electron emission angle along the $z$-polarization direction) as a function of $p_x$ at forward $p_x>0$ and backward $p_x<0$ electron emission directions. The fundamental harmonic field is linearly polarized along the $z$ axis and its peak intensity is fixed at 2$\times$ 10$^{14}$ W/cm$^2$ and the angular frequency is $\omega$=8.58 eV ($\lambda_1$=144.53 nm). The relative intensity ratio is fixed at $\eta=$0.63 and the he angular frequency is $\omega$=17.16 eV ($\lambda_2=\lambda_1/2$=72.26 nm). The results are shown for three relative optical phases: (a) $\delta\varphi$= $\pi/4$ (black curve); (b) $\delta\varphi$= $\pi/2$ (green curve); (c) $\delta\varphi$= $\pi$ (cyan curve). Also shown the spectrum (red dashed-line curve) in the absence of the streaking field. Inset: Angle-integrated photoelectron spectrum. The curves for different optical phases are indistinguishable.}
\end{figure*}

Taking advantage of these quantum-paths involved in one-photon and two-photon absorption processes, one can exploit their interferences to control the electron emission direction, and hence get insights into the continuum wave packets. This can be achieve by controlling the relative phase of these pathways through the variation of the relative optical phase. In general, the generated continuum wave packet is presumed to instantly follow the temporal evolution of the laser field. This suggests that the optical phase, which determines the timing of the field oscillations with respect to the pulse peak can be used to tackle the outgoing wave packet. This is true since the continuum wave packet born at the time $t$ will born with an offset at the time $t + \tau$, where $\tau$ = $\delta\phi/\omega_1$. Thus, the control of the ionization asymmetry can be achieved through the change of the relative optical phase $\delta\phi$. 

Indeed, changing the optical phase leads to a time-varying ionization of the electron owing to an offset in time of the birth of the continuum wave packet \cite{Agueny:18}. Therefore, electrons will be generated with different final momenta depending on the optical phase. The change of the final momentum can be described by the classical equation of motion $\Delta
\vec{p}(t_{ionz})=\int_{t_{ionz}}^\infty E(t)dt$, where $t_{ionz}$
is the ionization time. The final electron's momentum
$\vec{p}_{f}(t_{ionz})$ is then given by
 \begin{equation}\label{eq5}
\vec{p}_{f}(t_{ionz}) = \vec{p}_i + \Delta \vec{p}(t_{ionz}) =
\vec{p}_i + \vec{A(t_{ionz})},
 \end{equation}
where $\vec{p}_i$ is the initial momentum. It is clear that the
change on the final momentum is directly related to the vector
potential, and hence preserves the temporal information of the
electric field. Accordingly, by varying the optical phase one
can directly control the final momentum and hence get insight
into the ionization process. Information about this process can be revealed by the asymmetric distribution and its investigation helps to track the electron dynamics. 

The optical phase dependence upon the momentum distribution is shown in Fig. ~\ref{fig5}. Here calculations are performed at the peak intensity of the fundamental harmonic 2 $\times$10$^{14}$ W/cm$^2$, and for the relative intensity ratio $\eta=$0.63 (peak intensity of the streaking field is 8 $\times$10$^{13}$ W/cm$^2$). The results are shown for three relative optical phases: $\delta\varphi$= $\pi/4$, $\pi/2$, and $\pi$. It is seen that the asymmetric distribution is very sensitive to the change of the relative optical phase. In particular, the Autler-Townes doublet located along the $x$-polarization direction is found to be affected by this change, in contrast to the dynamic interference, which is seen to be insensitive to the optical phase. This sensitivity is imprinted in the transversal photoelectron spectrum [cf. Fig. ~\ref{fig5}(d)], which is shown in the backward emission direction, and is found to manifest by the broadening and enhancement of the Autler-Townes doublet. On the other hand, integrating the distribution over the electron emission angles washes out the quantum-path interferences, as shown in the inset of Fig. ~\ref{fig5}(d), where the curves for different optical phases are indistinguishable. For reference, the photoelectron spectrum obtained within the one-color scheme is also shown in the inset of Fig. ~\ref{fig5}(d) (red dashed-lines curve). Here, the signature of dynamic interference remains unchanged when introducing the control pulse, while the Autler-Townes doublet shows a slight enhancement of the ionization. 

The modification of the ionization asymmetry in the momentum distribution is a footprint of the ultrafast coherent control of the electron dynamics. In particular, changing the relative optical phase by $\pi$ results in a broader and larger peak of the Autler-Townes doublet, which is evidence that ionization rate and yield are enhanced. Thus, by manipulating the relative optical phase within the OTC scheme one can precisely tailor the optical cycles to yield to an unprecedented degree of control for the ultrafast ionization dynamics. On the other hand, the observed sensitivity to the optical phase demonstrates that the Autler-Townes effect results from electrons that are driven by the instantaneous electric field, while dynamic interference arises from those that follow the pulse envelope. Thus confirming that these two effects can be disentangled, and revealing the role of the OTC scheme for separating dynamical signals in the momentum distribution. Therefore, the ability of characterizing two dynamical effects occurring at the same time scale makes the OTC scheme ideal for probing the electron dynamics and characterizing the generation of the electron wavepacket and attosecond light pulses. Moreover, resolving dynamical signals by changing the optical phase within an OTC in a multi-cycle regime is shown to be efficient by carrying out angle-resolved photoelectron spectra. These are found to be more powerful for investigating coherent control mechanisms than the angle-integrated ones.

\section*{Discussion}


In conclusion, ultrafast ionization dynamics induced by OTC laser pulses has been theoretically studied
by solving the 2D-TDSE. The dynamics was shown to lead to the emergence of interference patterns in the photoelectron momentum distribution. We have identified the physical mechanism behind the observed pattern, and found to arise from the interplay between the dynamic interference caused by interference between emitted electrons in the rising and falling edges of the pulse envelope; and  quantum-path interferences between electron wave packet generated by one-photon and two-photon processes. These results were supported by additional calculations based on a coherent superposition model. Taking advantage of the combined OTC laser fields, we have characterized signals of the Autler-Townes effect and dynamic interference in momentum space through the control of spatial interference. Furthermore, we have demonstrated that the sensitivity of the symmetry-breaking of the momentum distribution to the change of the properties of the control scheme is a footprint of the ultrafast coherent control of the electron dynamics. In particular, it was shown that, by tuning the amplitude and the optical phase of the control pulse, the ionization rate can be controlled efficiently through the Autler-Townes effect. This was demonstrated to lead to enhancement of the ionization rate as a result of constructive interference between partial photoelectron waves having opposite-parity.

Our study provides a comprehensive picture of the crucial role of the OTC scheme to control the Autler-Townes effect and to characterize dynamical signals by carrying out the momentum distribution. In particular, the sensitivity of quantum-path interferences to various characteristics of the streaking field, and the ionization enhancement through the Autler-Townes effect can be exploited for accurate characterization and generation of isolated attosecond pulses.

\section*{Methods}

The TDSE for a hydrogen atom interacting with orthogonally polarized laser fields is expressed in two-dimensional (2D) cartesian coordinates. The 2D-model has been used in many theoretical studies (see for instance \cite{Xie:15,ZhangC:17}) and has been shown to reproduce well the full three-dimensional calculations (see the Supplemental Material for details about the calculations \cite{comment}), allowing thus a correct description of the dynamics involved in the hydrogen atom with less computational efforts. On the basis of this model, the TDSE is written as
\begin{equation}\label{eq1}
\Big[ H_0 + H_I(t) - i\frac{\partial}{\partial t}
\big]\psi(\vec{r},t)=0,
\end{equation}
where $\vec{r}=(x,z)$ denotes the vector position of the electron.
$H_0$ is the field-free Hamiltonian with a regularized 2D Coulomb potential \cite{Protopapas:97}
\begin{equation}\label{eq2}
H_0=-\frac{1}{2}\frac{\partial^{2}}{\partial z^{2}} 
-\frac{1}{2}\frac{\partial^{2}}{\partial x^{2}} -\frac{1}{\sqrt{x^2+z^2 + 0.64}}.
\end{equation}
The time-dependent interaction $H_I(t)$ is assumed to be linearly polarized along the $z$ axis for the fundamental harmonic field and along the $x$ axis for its second harmonic. This interaction is treated in the velocity gauge and can be expressed within the dipole interaction as
\begin{equation}\label{eq3}
H_I(t)= -iE_{0}\Bigg( \frac{g_1(t)}{\omega_1}\sin(\omega_1 t) \vec{e}_z + 
\eta\frac{g_2(t)}{\omega_2}\sin(\omega_2 t  + \delta\phi)\vec{e}_x \Bigg)\cdot\vec{\nabla}
\end{equation}
where  $\omega_1$ and $\omega_2$ represent the angular frequency of the fundamental harmonic field ($E_{0}$ is its maximum field strength) and its second harmonic ($E_{1}=\eta E_{0}$ is its maximum field strength), respectively. Here, $\eta=\sqrt{I_1/I_0}$ and $\delta\phi$ represent the relative ratio of the peak intensity (defined as $I_i=E_i^2$, $(i=0,1)$) and the relative carrier envelope phase of the OTC laser pulses, respectively.
$g_i(t)$ (i=1,2) is the pulse envelope which has a cosinus-shaped $\cos^2(\pi t/\tau_i)$ form with $\tau_i=2\pi N_c/\omega_i$ is the total duration of the pulse. $N_c$ is the total number of cycles. The parameter $\eta$ determines the role of the second pulse in the OTC scheme \cite{Zhang:14a,Richter:15,Geng:15}: in the case of $\eta< 1$, the second pulse does not contribute significantly to the ionization process, but rather acts as a control pulse; while for $\eta \sim 1$ the two pulses contribute significantly to the ionization of the electron and even can compete. 

We consider the hydrogen atom to be in the ground state. The time evolution of the electronic wave function
$\psi(x,z)$, which satisfies the TDSE [cf. Eq. (\ref{eq1})], is solved
numerically using a split-operator method combined with a fast Fourier
transform algorithm. This is carried out on a symmetric grid
of size $|z|$ = $|x|$=1024 a.u. with the spacing grid
$\delta z$ = $\delta x$ =0.25 a.u., i.e. 4096 grid
points along $z$- and $x$-axis directions. 

The time step used in the simulation is $\delta t=0.05$ a.u.. The convergence
is checked by performing additional calculations with twice the size
of the box and a smaller time step. An absorbing boundary is employed to avoid artificial reflections, but without
perturbing the inner part of the wave function. The boundary is
chosen to span 10\% of the grid size in each direction. At the end
of the interaction $t=t_f$, we calculate the 2D-
momentum distribution of the photoelectron from the Fourier
transform of the spatial ionization wave function $\psi_{ioniz}$ \cite{Agueny:16a,Agueny:16b}. The latter is evaluated using a projection technique \cite{Kukulin:78,Agueny:15}. In this technique, we consider orthogonal projection operators $\hat{P}$ and $\hat{Q}$, such that
\begin{equation}\label{PQ}
\hat{Q}=1-\hat{P}=1 - \sum_n^N  | \psi_n \rangle \langle  \psi_n | ,
\end{equation}
where $| \psi_n \rangle $ are the eigenvectors of the field-free Hamiltonian $H_0$ [cf. Eq. (\ref{eq2})]. The spatial ionization wave function $\psi_{ioniz}$ can be written 
\begin{equation}\label{PQioniz}
| \psi_{ioniz} \rangle = \hat{Q} | \psi(t_f) \rangle.
\end{equation}
Substituting the equation (\ref{PQ}) into the equation (\ref{PQioniz}) we obtain $\psi_{ioniz}$
\begin{equation}\label{ioniz}
\psi_{ioniz}(\vec{r})= \psi(\vec{r},t_f) - \sum_n \psi_n(\vec{r}) \int \psi_n(\vec{r}')\psi(\vec{r}',t_f) d\vec{r}',
\end{equation}
where the sum over $n$ covers the first nine bound states, which are obtained using the imaginary time
propagation. The convergence of the ionization wave function with respect to the projection time $t_f$ has been checked. This is done by continuing propagating the wave packet after the pulse is turned off for an additional 150 a.u.. The momentum distributions obtained from the square of a Fourier transform of the spatial ionization wave function [cf. Eq. (\ref{ioniz})] are found to be unchanged, ensuring the convergence of the results presented in this work. 

For the calculation of the momentum distribution in the rising and falling edges of the pulse envelope, we adopte the same numerical procedure as described by equation (\ref{ioniz}). In the rising part of the half-pulse, the the time-dependent wave function is collected at the time corresponding to the maximum of the pulse envelope, and the sum in equation (\ref{ioniz}) covers the dressed bound states. The obtained ionization wavefunction is labeled $\psi_{rising}$. In the same way we obtain the ionization wave function in the falling part of the half-pulse, which we denote $\psi_{falling}$. Here the propagation scheme starts at the maximum of the pulse envelope and the initial state is the dressed ground state. A coherent sum of these two wavefunctions (i.e. $|\psi_{rising} + \psi_{falling}|^2$) is expected  to induce quantum interferences between electrons generated in the rising part and those in the falling part of the pulse envelope as shown in Fig.\ref{fig2}(b).

\section*{Data Availability Statement}
The datasets generated during the current study are available from the corresponding author on reasonable request.






 
\section*{Additional Information}
{\bf Competing Interests:} The author declares no competing interests.








\end{document}